\documentstyle[psfig]{aipproc}

\def\etal{{\it et al.}}

\begin{document}

\begin{flushright}
hep-ph/0012340 \\
UCCHEP/15-00
\end{flushright}
\vskip -.7cm

\title{Radiative Corrections to Chargino Production with
Polarized Beams\footnote{Talk given at Linear Collider Workshop 
2000--LCWS, Fermilab, Chicago, October 24-28, 2000.}}

\author{{\underline{Marco A. D\'\i az}}$^{1,4}$, Steve F. King$^2$, and 
Douglas A. Ross$^{2,3}$}
\address{
$^1$Departamento de F\'\i sica, Universidad Cat\'olica de Chile,
Macul, Santiago 6904411, Chile
\\
$^2$Department of Physics and Astro., University of Southampton,
Southampton SO17 1BJ, U.K.
\\
$^3$Theory Division, CERN, Geneva, Switzerland
\\
$^4$Abdus Salam ICTP, Strada Costiera 11, 34100 Trieste, Italy
}

%\lefthead{LEFT head}
%\righthead{RIGHT head}
\maketitle

\vskip -.6cm
\begin{abstract}

We show that radiative corrections to chargino production in 
electron--positron annihilation with polarized beams can be large
especially in the case of right handed electrons. In addition, 
there is some dependency on the squark masses that allows us to
extract information about the squark spectrum from the chargino
production.

\end{abstract}

If supersymmetry is realized in nature, charginos\footnote{
See \cite{exp} for latest experimental (negative) results on chargino 
searches.
}
should be discovered 
at the LHC (or at the Tevatron if they are light enough). Nevertheless,
it is at a future electron-positron collider where the necessary precision
\cite{MB} on the measurements of masses and cross-sections can be achieved 
in order to extract the fundamental SUSY parameters of the theory
\cite{unpolar}. Furthermore, knowing the SUSY parameters at the weak 
scale, allow us to infer information on physical principles at very high 
energy scales \cite{BDQT}. In this way, it is clear the importance of 
including one-loop radiative corrections to the masses \cite{PP} and 
cross sections \cite{DKR,othUnp}.

The importance of beam polarization and chargino helicities 
\cite{polarized} has also been stressed, and radiative corrections to 
chargino production with polarized electron and positron beams have been 
calculated recently \cite{DKR1,BH}. Here we report on this calculation
\cite{DKR1}, and show that most of the time the corrections are 
non-negligible, and some times they are very large, especially when the 
electrons are right-handed polarized. The calculation is done in the 
$\overline{MS}$ scheme and we include quarks and squarks in the loops, 
gaining a color factor $N_c=3$ over the rest of the loops and an 
additional factor of $N_g=3$ over loops involving gauge and Higgs bosons.
Charginos can be produced in electron-positron annihilation via 
intermediate $Z$ and photons in the s-channel and via electron-sneutrinos 
in the t-channel. The corrections reported here can be organize into
$Z\tilde\chi^+\tilde\chi^-$, $\gamma\tilde\chi^+\tilde\chi^-$, and
$e^\pm\tilde\nu_e\tilde\chi^\mp$ corrected vertices \cite{DKR1}.

\begin{figure}
\centerline{\protect\hbox{\psfig{file=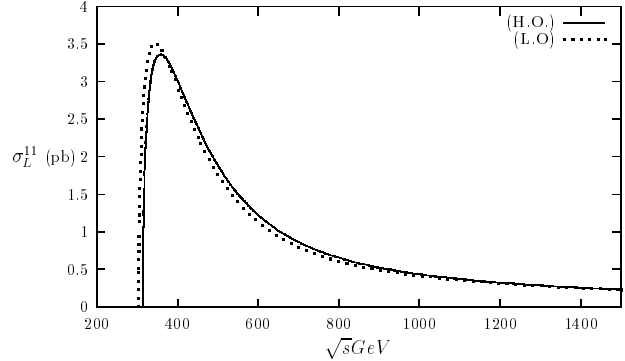,width=0.55
\textwidth}}}
\caption{Lowest order (L.O.) and higher order (H.O.)
cross-section for lightest chargino pair production
for left-polarized electrons with $\tan \beta =5$ and the
other parameters as given in the text.} 
\label{L5} 
\end{figure} 
Fig.~\ref{L5} shows the cross-section for a beam of left-handed electrons
for the production of two lightest charginos. We consider running parameters
$M_2=165$ GeV, $\mu=400$ GeV, and $\tan\beta=5$ at the scale $Q=m_z$. For 
the squark soft parameters we take, for simplicity, degenerate values 
$M_Q=M_U=M_D=500$ GeV and $A\equiv A_U=A_D=500$ GeV. For definiteness, we 
take the sneutrino mass to be degenerate with the squark mass parameters. 
We observe that the peak of the cross section is reduced due to one-loop 
radiative corrections by $4\%$. The values of the chargino masses in this 
case are: 151 and 421 GeV for the running masses and 157 and 431 GeV for 
the pole masses. Therefore, the shift in the threshold in Fig.~\ref{L5} 
corresponds to an increase of $4\%$ in the lightest chargino mass.

\begin{figure}
\centerline{\protect\hbox{\psfig{file=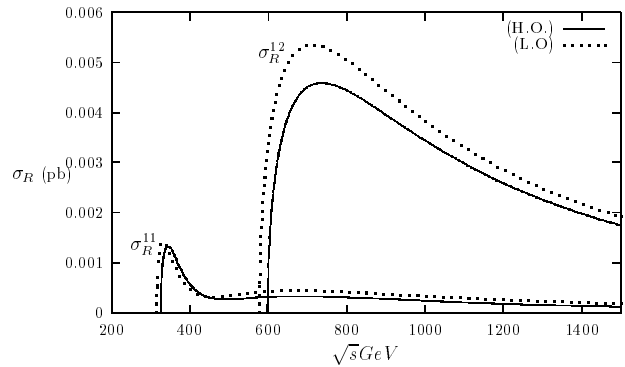,width=0.55
\textwidth}}}
\caption{Lowest order (L.O.) and higher order (H.O.)
cross-sections for lightest and unequal mass chargino
pair production for
right-polarized electrons with $\tan \beta =50$ and the
other parameters as given in the text.} 
\label{R50} 
\end{figure} 
Cross-sections for a beam with right-handed electrons are smaller than 
for the left-handed electron case. For this reason, the effect of 
radiative corrections tends to be more important. In addition, at large
values of $\tan\beta$ radiative corrections are also larger due to the
increase on the top and bottom yukawa couplings. This can be seen in 
Fig.~\ref{R50} where we have taken $\tan\beta=50$ \footnote{
This high value of $\tan\beta$ is favored by $SO(10)$ inspired models 
with top-bottom-tau Yukawa unification and consistent radiative 
electroweak symmetry breaking, obtained after the inclusion of 
D-terms\cite{so10,BDQT}.
}, 
with the rest of the parameters as in the previous figure. Corrections to
the peak cross section are $-14\%$ for the case of mixed production of
$\tilde\chi^+_1\tilde\chi^-_2$ (a factor of two should be added if we 
include the case $\tilde\chi^+_2\tilde\chi^-_1$), and $-6\%$ for the 
production of a pair of the lightest charginos. Despite the smallness 
of the cross-section, with luminosities of the order of $1\,ab^{-1}$
these signals will be measurable. Note that right-handed electrons do 
not couple to charginos and, therefore, the sneutrino contribution to
the cross section $\sigma_R$ is not present.

\begin{figure}
\centerline{\protect\hbox
{\psfig{file=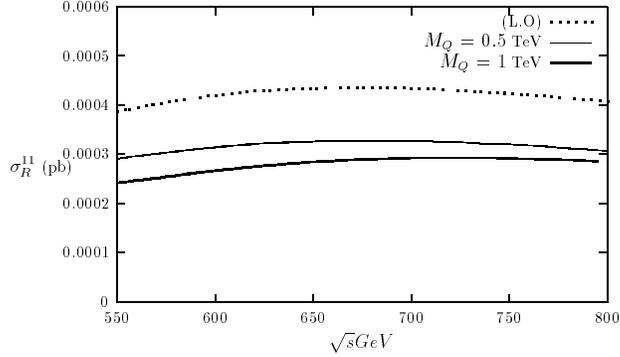,width=0.55
\textwidth}}}
\caption{Detailed blow-up of
cross-sections for right-polarized electrons
for the lightest chargino pair with $\tan \beta =50$.
The H.O. cross-sections are for degenerate squark soft mass
parameters of 0.5 TeV and 1 TeV.} 
\label{R50blowup11} 
\end{figure}

Corrections to $\sigma_R^{11}$ away from the peak of the cross-section are 
larger. In Fig.~\ref{R50blowup11} we show a blow-up of the region with 
intermediate values of the center of mass energy. For example, for
$\sqrt{s}=650$ GeV the one-loop corrected cross-section is $25\%$ smaller
than the tree level cross section. In addition, we show the case where
the squark mass parameters are degenerate and equal to 1 TeV. In this case,
for $\sqrt{s}=650$ GeV the correction is $-35\%$. This squark mass scale
dependency permits the extraction of information on the squark sector
from chargino observables.

\begin{figure}
\centerline{\protect\hbox
{\psfig{file=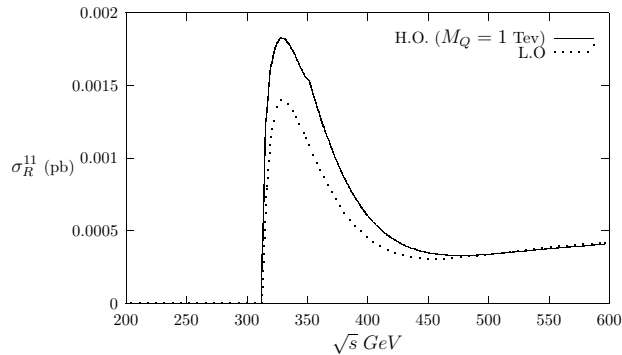,width=0.55
\textwidth}}}
\caption{L.O. and H.O. cross sections for $\tan\beta=50$ and degenerate 
squark soft mass parameters of 1 TeV, for the case where $\mu$ and $M_2$
are modified such that the chargino pole masses remain the same.
} 
\label{fixchm2} 
\end{figure}

The large radiative corrections to the chargino production cross
section observed in the previous figures cannot be explained simply by 
a shift in the chargino masses\cite{DKR1}. This can be seen in 
Fig.~\ref{fixchm2} where we compare the corrected cross-section with
a ``modified tree-level'' cross-section calculated with running
masses whose numerical value are the same as the renormalized masses
(therefore, calculated with a different set of values for $M_2$ and 
$\mu$). In this way, in Fig.~\ref{fixchm2} we see the genuine corrections
to the cross section, which corresponds to a $31\%$ increase of the peak 
of the cross-section for 1 TeV squark mass parameters ($20\%$ if the 
squark mass parameters are equal to 500 GeV).

In summary, we have shown that it is crucial to include the one-loop 
radiative corrections to the chargino masses and cross sections in
order to correctly extract the fundamental SUSY parameters from the 
experimental observables. Especially large corrections are obtained 
at large values of $\tan\beta$ and for right handed polarized electrons.
Since dependence exist on the squark masses, information on the squak
spectrum can be inferred from the chargino observables.

\end{document}